\documentclass[superscriptaddress,prb,twocolumn,footinbib]{revtex4-2}

\usepackage[bookmarks=false,colorlinks=true,urlcolor=blue,citecolor=blue,linkcolor=blue]{hyperref}

\usepackage{tikz}
\usetikzlibrary{calc,arrows}

\usepackage{amsmath}
\usepackage{gensymb}

\usepackage{booktabs}

\usepackage[compat=1.1.0]{tikz-feynman}
\tikzfeynmanset{/tikzfeynman/warn luatex=false}

\newcommand{\mycomment}[1]{}

\newcommand{\markup}[1]{#1}

\newcommand{\be}{\begin{equation}}
\newcommand{\ee}{\end{equation}}
\newcommand{\bea}{\begin{eqnarray}}
\newcommand{\eea}{\end{eqnarray}}
\newcommand{\beal}{\begin{align}}
\newcommand{\eeal}{\end{align}}
\newcommand{\bes}{\begin{equation} \begin{split}}
\newcommand{\ees}{\end{split} \end{equation}}

\newcommand{\qv}{\vec{q}}

\newcommand{\Qv}{\vec{Q}}

\newcommand{\rv}{\vec{r}}
\newcommand{\rvp}{\vec{r}^{\,\prime}}
\newcommand{\Rv}{\vec{R}}

\newcommand{\Sv}{\vec{S}}

\newcommand{\bv}{\vec{b}}

\newcommand{\Kinv}{\ensuremath{K^{-1}}}
\newcommand{\Keff}{K}
\newcommand{\Keffinv}{K^{-1}}

\newcommand{\RN}[1]{%
\textup{\uppercase\expandafter{\romannumeral#1}}%
}

\newcommand{\Tr}{\mathrm{Tr}}

\usepackage{contour}

\definecolor{taylorswift}{rgb}{0.0862745098,0.4666666667,0.3411764706}
\definecolor{fearless}{rgb}{0.8862745098,0.6117647059,0.2823529412}
\definecolor{speaknow}{rgb}{0.4588235294,0.2274509804,0.4980392157}
\definecolor{red}{rgb}{0.6509803922,0.1254901961,0.2705882353}
\definecolor{TS1989}{rgb}{0.1803921569,0.6,0.9764705882}
\definecolor{reputation}{rgb}{0.1450980392,0.1490196078,0.1529411765}
\definecolor{lover}{rgb}{0.8392156863,0.2117647059,0.5529411765}

\begin{document}

\title{Sublattice Pairing in Pyrochlore Heisenberg Antiferromagnets}

\author{Cecilie Glittum}
\affiliation{T.C.M. Group, Cavendish Laboratory, JJ Thomson Avenue, Cambridge CB3 0HE, United Kingdom}
\affiliation{Department of Physics, University of Oslo, P.~O.~Box 1048 Blindern, N-0316 Oslo, Norway}
\author{Olav F. Sylju{\aa}sen}
\affiliation{Department of Physics, University of Oslo, P.~O.~Box 1048 Blindern, N-0316 Oslo, Norway}

\begin{abstract}
We argue that classical pyrochlore Heisenberg antiferromagnets with small further-neighbor couplings can order in a state where pairs of sublattices form antiparallel spirals. The spiral ordering wave vectors of the two pairs are in general different from each other, and are constrained by which sublattices are being paired. This sublattice pairing state generally breaks inversion and most rotation symmetries. Its existence depends on the antiferromagnetic nearest-neighbor coupling which favors the spins on each tetrahedron to sum to zero. To substantiate our argument, we extend the nematic bond theory; a diagrammatic large-$N_s$ method, to non-Bravais lattices, and we demonstrate that the predicted state is indeed realized at low temperatures in a large region of exchange coupling space. We also carry out a spin wave calculation which suggests that the sublattice pairing state is coplanar.
\end{abstract}

\date{\today}

\maketitle

\section{Introduction \label{sec:introduction}}
\noindent

The Heisenberg antiferromagnet on the pyrochlore lattice has gotten much attention as it is a spin liquid candidate.
This is mainly motivated by the antiferromagnetic (AF) nearest-neighbor classical model, which is predicted to be disordered at all temperatures~\cite{Villain1979,Reimers91,Reimers92,Moessner98}. However, real pyrochlore magnetic materials are seldom described by the nearest-neighbor model alone. It is therefore important to understand the effects of further-neighbor couplings, and when and what magnetic order they may cause.

In this article we propose a new kind of ordered state for pyrochlore Heisenberg antiferromagnets: a sublattice pairing (SLP) state,  where sublattices pair up, and each pair form antiparallel spirals.

Ordering transitions as a result of adding further-neighbor couplings has been studied in mean-field theory~\cite{Reimers91}, and it is known that further-neighbor interactions induce symmetry breaking in the purely classical $J_1$-$J_2$ model~\cite{Nakamura07,Tsuneishi07,Chern08,Okubo2011}.

\begin{figure}
\includegraphics[width=\columnwidth]{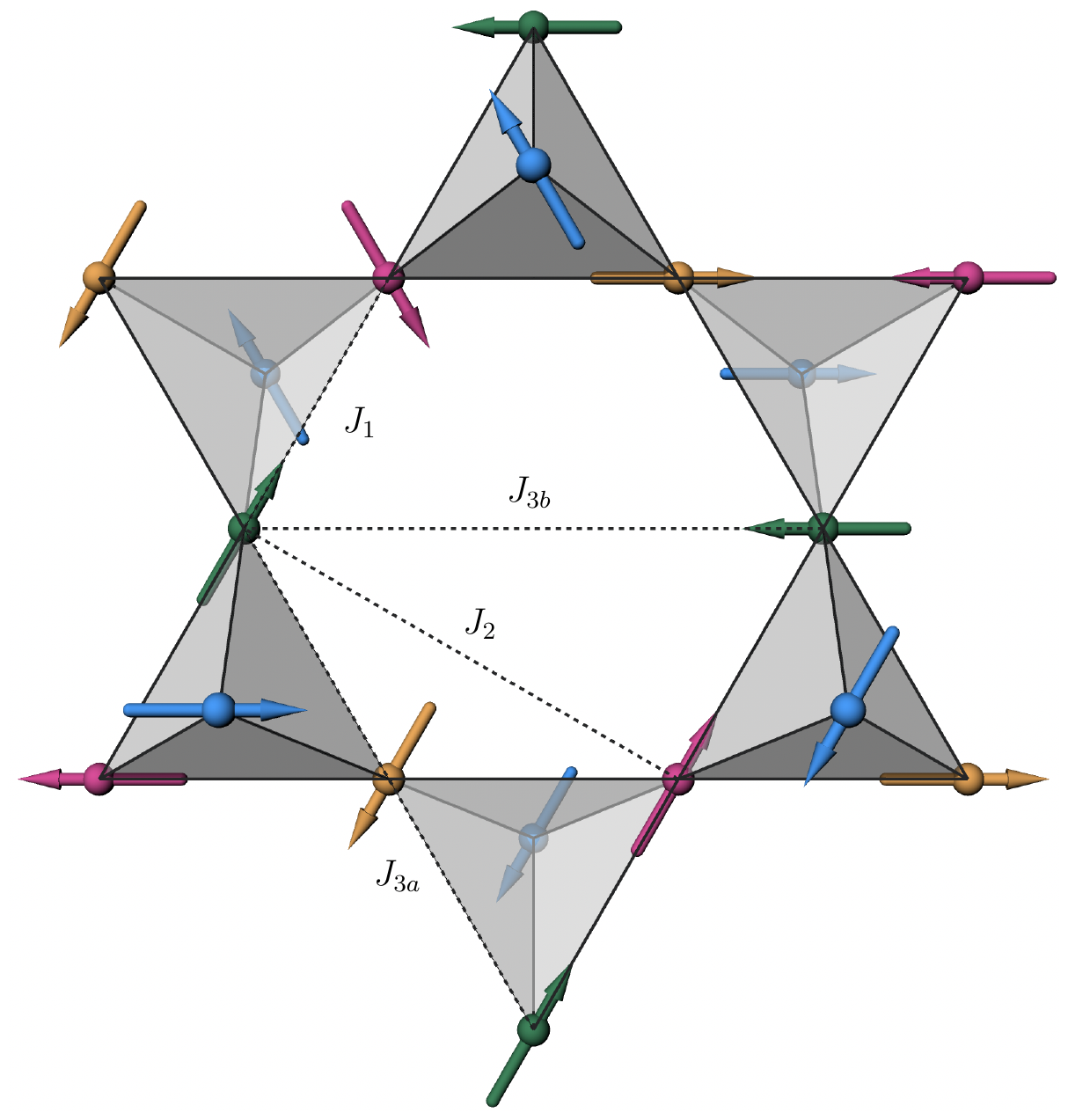}
\caption{Pyrochlore lattice with spins showing an example of an SLP state where sublattices 0 (pink) and 1 (blue) are antiparallel and ordered at $\vec{Q}_{(0,1)} = (0, -4\pi/3, 4\pi/3)$ and sublattices 2 (green) and 3 (yellow) are antiparallel and ordered at $\vec{Q}_{(2,3)} = (0, 4\pi/3, 4\pi/3)$. The first ($J_1$), the second ($J_2$) and the two inequivalent third ($J_{3a}$ and $J_{3b}$) nearest-neighbor couplings are shown. The blue spins show how layers of triangular planes separated by kagome layers of the remaining sublattices are ordered in $120\degree$ spirals for the $J_1$-$J_{3b}$ model.
\label{fig:120config}}
\end{figure}

The third nearest-neighbor couplings are known to be important for several pyrochlore materials, and in many cases more important than the second nearest-neighbor couplings~\cite{Wills2006,Yaresko08,Cheng08}. There are two inequivalent third nearest-neighbor couplings on the pyrochlore lattice: $J_{3a}$ which goes in the direction of $J_1$,
and $J_{3b}$, which goes across the hexagons, see Fig.~\ref{fig:120config}.
Existing theoretical works including third nearest-neighbor couplings either treat the two as equal or set $J_{3b}$ to zero when studying ordering transitions~\citep{Nakamura07,ConlonChalker2010, Mizoguchi2018}.

In this article we treat the Heisenberg antiferromagnets classically. We focus on the hierarchy of magnetic scales $J_1 > J_{3b} \geq J_2, J_{3a}$, which might be important for the $\mathrm{Gd}_2B_2\mathrm{O}_7$ pyrochlores~\cite{Wills2006} and the $A\mathrm{Fe}_2\mathrm{O}_4$ spinels~\cite{Cheng08}.

The Hamiltonian is
\be \label{eq:Hamiltonian}
H = \frac{1}{2} \sum_{\rv, \rvp} J(\rvp - \rv) \; \Sv_{\rv}\cdot \Sv_{\rvp},
\ee
where the exchange couplings are illustrated in Fig.~\ref{fig:120config}. The pyrochlore lattice has four fcc sublattices. We label the spins by their unit cell $\Rv$ and sublattice index $i$ rather than position $\rv = \Rv + \vec{\alpha}_i$. $\Rv$ is constructed from the fcc primitive lattice vectors $\vec{a}_1 = (0,1/2,1/2), \vec{a}_2 = (1/2,0,1/2)$,  and $\vec{a}_3 = (1/2,1/2,0)$, where we have set the cubic lattice constant to unity. The sublattice vectors $\vec{\alpha}_i$ are $\vec{\alpha}_0=(0,0,0)$ and $\vec{\alpha}_{i} = \vec{a}_i/2$ for $i=\{1,2,3 \}$. We choose energy units $J_1=1$.

 \section{Sublattice pairing} \label{sec:sublatticepairing}

\begin{figure}[]
\includegraphics[width=\columnwidth]{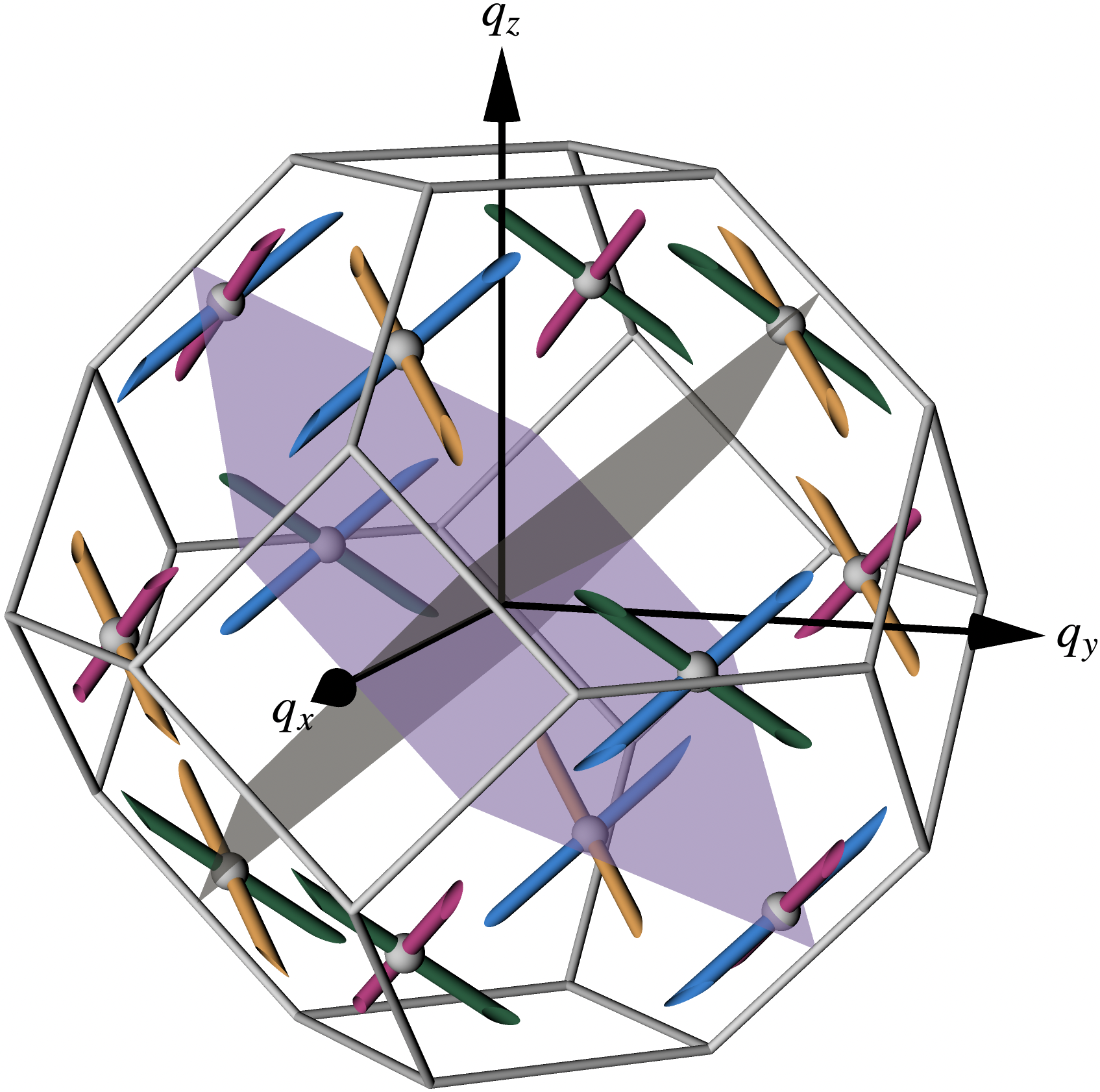}
\caption{The first Brillouin zone (1BZ) showing planes $\Qv_{(0,1)}$ (purple) and $\Qv_{(2,3)}$ (grey) satisfying Eq.~\eqref{eq:down-tetrahedra-condition}, and $J_{3b}$ energy minimal lines for sublattices $0$ (pink), $1$ (blue), $2$ (green), and $3$ (yellow). The lines intersect at $(0,4\pi/3,4\pi/3)$ and symmetry-related points (light grey). The line intersection points in the planes correspond to the possible spiral ordering wave vectors for SLP between sublattices $(0,1)$ and $(2,3)$.} \label{fig:120rods}
\end{figure}

An AF nearest-neighbor coupling favors the spins on each tetrahedron to sum to zero~\cite{Reimers91, Villain1979}, i.e. $\sum_{i = 0}^3 S_{\Rv, i} = 0$ for the up-tetrahedra and $\sum_{i=0}^3 \Sv_{\Rv - \vec{a}_i, i} = 0$ for the down-tetrahedra.  If each sublattice orders in a single-$\qv$ spiral state $S_{\Rv, i} = \vec{u}_i\cos \Qv_i\cdot \Rv + \vec{v}_i \sin\Qv_i\cdot\Rv$, the Fourier-transformed condition for the up-tetrahedra gives
\begin{equation}
\sum_i \left[ \left(\vec{u}_i - i\vec{v}_i \right) \delta_{\qv,\Qv_i} + \left( \vec{u}_i + i \vec{v}_i \right) \delta_{\qv,-\Qv_i}\right] = 0,
\end{equation}
where the $\delta$'s should be understood modulo a reciprocal lattice vector.
This is satisfied by what we refer to as an SLP state. In an SLP state the sublattices form pairs, such that each sublattice pair $(i,j)$ shares the same ordering wave vector $\Qv_{(i,j)}$ and has antiparallel spins:
\begin{align}\label{eq:genstate}
\Sv_{\Rv, i} &= \vec{u}_{(i,j)} \cos (\Qv_{(i,j)} \cdot \Rv) + \vec{v}_{(i,j)} \sin (\Qv_{(i,j)} \cdot \Rv),\\
\Sv_{\Rv, j} &= -\Sv_{\Rv, i},
\end{align}
where $\vec{u}_{(i,j)}$ and $\vec{v}_{(i,j)}$ are orthonormal vectors.
If the two ordering wave vectors are different, this state satisfies also the condition for the down-tetrahedra if
\be\label{eq:down-tetrahedra-condition}
\left[\Qv_{(i,j)} \cdot (\vec{a}_i - \vec{a}_j)\right]\mod 2\pi = 0
\ee
for both pairs of sublattices. Figure~\ref{fig:120rods} shows the planes in momentum space where $\Qv_{(0,1)}$ and $\Qv_{(2,3)}$ satisfy this equation.

The ordering wave vectors $\Qv_{(i,j)}$ are generally found by minimizing the energy. As the SLP state minimizes the $J_1$ terms of the energy, it is sufficient to minimize the further-neighbor energy terms subject to the condition Eq.~\eqref{eq:down-tetrahedra-condition}.
As a first example, we consider the pure $J_1$-$J_{3b}$ model. The third nearest neighbors on the pyrochlore lattice couple sites from the same fcc sublattices, and $J_{3b}$ alone effectively reduces each of the four fcc sublattices to a set of decoupled parallel triangular planes.
An AF $J_{3b}$ will then favor $120\degree$ order in each triangular plane. For a single plane, there are two such chiral ordering vectors, given by clockwise and counter-clockwise $120\degree$ rotations. Since the triangular planes in a set are decoupled, the addition of any wave vector orthogonal to the triangular planes will still give $120\degree$ order in each plane, but with an additional inter-plane rotation.
This gives rise to a set of lines in momentum space for each of the four sublattices, along which the $J_{3b}$-part of the energy is minimal. This is illustrated in Fig.~\ref{fig:120rods}. These lines would correspond to rods of scattering if $J_1=0$.
The lines intersect at points where the $J_{3b}$ energy of two sublattices is minimized by the same $\Qv$ vector. These wave vectors are given by $(0,4\pi/3, 4\pi/3)$ and symmetry-related vectors
and lie also on the planes satisfying the tetrahedron condition Eq.~\eqref{eq:down-tetrahedra-condition}.  The example shown in Fig.~\ref{fig:120rods} is $\vec{Q}_{(0,1)}=\pm(0,4\pi/3,-4\pi/3)$ and $\vec{Q}_{(2,3)}=\pm(0,4\pi/3,4\pi/3)$, and the corresponding SLP $120\degree$  configuration is illustrated in Fig.~\ref{fig:120config}.

To  study the ordering wave vectors for the more general  $J_1$-$J_2$-$J_{3a}$-$J_{3b}$ model, we make use of the fact that this model can be recast into a  $\tilde{J}_1$-$\tilde{J}_{3a}$-$J_{3b}$ model when the tetrahedra conditions are satisfied, with $\tilde{J}_1 = J_1 - J_2$ and $\tilde{J}_{3a}=J_{3a} - J_2$~\cite{Chern08}. $\Qv_{(i,j)}$ is then found as the wave vector that satisfies Eq.~\eqref{eq:down-tetrahedra-condition} and minimizes the $\tilde{J}_{3a}$-$J_{3b}$ energy sum of the pairing fcc sublattices $i$ and $j$.  For $-3 \leq \tilde{J}_{3a}/J_{3b} \leq 1$ with AF $J_{3b}$, we find that $\Qv_{(i,j)}$ is given by the vectors symmetry related to
\begin{equation}\label{eq:SLPQs}
\Qv = \begin{cases}
        (0,h,h),  & \tilde{J}_{3a}/J_{3b} \leq \sqrt{2}-1 \\
       (2\pi, h-2\pi, h-2\pi),  \!\!\! & \tilde{J}_{3a}/J_{3b} >  \sqrt{2}-1,\\
     \end{cases}
\end{equation}
with $h \!\! = \!\! 2\arccos\left[-(1+\tilde{J}_{3a}/J_{3b})/2\right]$, that satisfy Eq.~\eqref{eq:down-tetrahedra-condition}.

When $\tilde{J}_{3a}/J_{3b} < -3$ for AF $J_{3b}$, $\tilde{J}_{3a}/|J_{3b}| < 1$ for ferromagnetic (FM) $J_{3b}$, or $\tilde{J}_{3a} < 0$ for $J_{3b} = 0$, the minimum occurs at $\Gamma$. The associated SLP state, SLP-$\Gamma$, where both the ordering vectors are equal to $\Gamma$, covers both the collinear Néel state~\cite{Chern08} and the coplanar Palmer-Chalker state~\cite{PalmerChalker2000}. When $\tilde{J}_{3a} > |J_{3b}|$, the minimum occurs at $\mathrm{X_{\mathrm{1BZ}}}$.

For $\tilde{J}_{3a}/J_{3b} = -1$,  there is a line minimum: $\Qv = (l,\pi, \pi)$ for AF $J_{3b}$ and $\Qv = (l, 0, 0)$ for FM $J_{3b}$.

\section{Nematic Bond Theory \label{sec:method}}

In order to investigate the occurrence of SLP states in the pyrochlore Heisenberg model, Eq.~\eqref{eq:Hamiltonian}, we employ the nematic bond theory (NBT)~\cite{Schecter2017}.
The NBT is a large-$N_s$ approximation, where $N_s$ is the number of spin components, leading to a set of self-consistent equations for classical Heisenberg magnets.
It has previously been employed to the square, cubic and triangular lattices~\cite{Schecter2017, Syljuaasen2019, Glittum2021, Liu2022}. In this article, we extend the NBT to non-Bravais lattices with $m$ sublattices.
Consequently, quantities like the exchange coupling $J_{\qv}$ and the self-energy $\Sigma_{\qv}$ become $m\times m$ matrices in sublattice space.

In momentum space, the Hamiltonian is
\be
H = \sum_{\qv}\sum_{ij} J_{\qv, ij} \Sv_{-\qv, i} \cdot \Sv_{\qv, j},
\ee
where the $\qv$-sum goes over the first Brillouin zone, and
$i$ and $j$ are sublattice indices.
In the NBT, the classical spins are integrated out by introducing a constraint field $\lambda_{\Rv, i}$ ensuring unit length of the spins, $|\Sv_{\Rv, i} | = 1$, through
\be
\delta(|\Sv_{\Rv, i}|-1) = \int_{-\infty}^{\infty}\frac{\beta d\lambda_{\Rv, i}}{\pi} e^{-i\beta \lambda_{\Rv, i} (\Sv_{\Rv, i}\cdot \Sv_{\Rv, i}-1)}.
\ee

The remaining integrals over the constraint field are treated separately for the average constraints $\Delta_i$ and the fluctuations around the average. The average constraints are treated by the saddle-point approximation, and the fluctuations are treated through diagrammatic perturbation theory (large-$N_s$ approximation).
As in Ref.~\onlinecite{Schecter2017}, the diagrams consist of solid and wavy lines, representing spin and constraint propagators, respectively. In addition to sublattice indices, we have here also included directions to the spin propagators to allow for breaking of inversion symmetry.  The spin propagator $\Kinv_{\qv, ij}$ is then to be understood as carrying momentum $\qv$ from $i$ to $j$.

\begin{figure}[]
\includegraphics[width=\columnwidth]{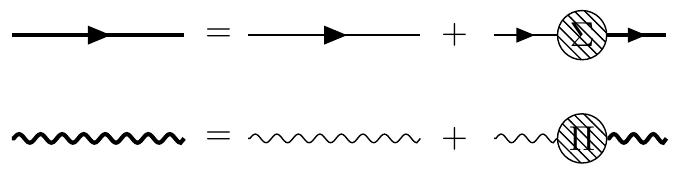}
\caption{Dyson equations for the renormalized spin propagator $\Keffinv_{\qv}$ (bold solid line), and the renormalized constraint propagator $D_{\qv}$ (bold wavy line). \label{Dyson}}
\end{figure}

\begin{figure}[]
\includegraphics[width=0.73\columnwidth]{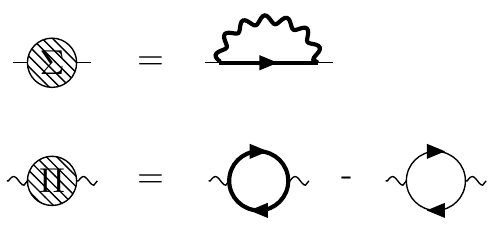}
\caption{Self-consistent equations for the self-energy $\Sigma_{\qv}$ and the polarization $\Pi_{\qv}$. \label{fig:selfconsistentdiagrams}}
\end{figure}

The spin and constraint propagators are renormalized by the self-energy $\Sigma_{\qv}$ and the polarization $\Pi_{\qv}$, respectively, through the Dyson equations in Fig.~\ref{Dyson}.
The large-$N_s$ approximation is performed through a set of self-consistent equations for the self-energy and the polarization, which are shown diagramatically in Fig.~\ref{fig:selfconsistentdiagrams}.  These equations approximate the self-energy and the polarization by infinite resummations of classes of diagrams excluding vertex corrections.
Combining the Dyson equations and the self-consistent equations, the NBT equations are~\cite{Schecter2017}
\begin{align}
&\Keff_{\qv, ij} = J_{\qv, ij} + \Delta_i\delta_{ij} - \Sigma_{\qv, ij},\\
&D^{-1}_{\qv, ij} = \frac{N_s}{2} \sum_{\vec{p}} \Keffinv_{\qv + \vec{p}, ij
}\Keffinv_{\vec{p}, ji}, \label{eq: NBT: Dinv self-consistent} \\
&\Sigma_{\qv, ij} = -\sum_{\vec{p} \neq 0}\Keffinv_{\qv - \vec{p}, ij}D_{\vec{p}, ij}, \label{eq: NBT: Sigma self-consistent}
\end{align}
where $\Keffinv_{\qv}$ is the renormalized spin propagator and $D_{\qv}$ is the renormalized constraint propagator.

As the constraint field now has a sublattice index, we get a separate saddle-point equation for each sublattice
\be
\frac{N_sT}{2V} \sum_{\qv} \Keffinv_{\qv, ii} = 1, \label{eq:T}
\ee
where $V=L^3$ is the number of unit cells. These $m$ saddle-point equations give the temperature \markup{$T$}. They must all give the same temperature for the solution to be physical. Following the derivation in Ref.~\onlinecite{Glittum2021}, the free energy per unit cell (excluding vertex corrections) is
\begin{equation}\label{eq: f}
\begin{split}
f = & -\sum_{i} \Delta_i + \frac{T}{2V}\sum_{\qv}\ln\det\left(\frac{T^2}{2V}D^{-1}_{\qv}\right) \\
&- \frac{N_sT}{2V} \sum_{\qv}\left[\ln\det\left(T\Keffinv_{\qv}\right) - \Tr\left(\Keffinv_{\qv}\; \Sigma_{\qv}\right) \right] \\
 & - m\frac{N_s-1}{2}T\ln\pi.
 \end{split}
\end{equation}

The self-consistent equations are solved by iteration starting from a random self-energy and equal values of the $\Delta_i$s. Each iteration gives an overall negative contribution to the self-energy. To avoid the general increase in temperature associated with this, the $\Delta_i$s are renormalized in every iteration by subtracting from them the minimum eigenvalue among all $\Sigma_{\qv}$. In addition\markup{,} each $\Delta_i$ is adjusted very slightly so that Eq.~\eqref{eq:T} gives the same value of the temperature for all sublattices. We iterate until the temperature has converged, and then employ
$\Keffinv_{\qv}$, $\Sigma_{\qv}$ and $D_{\qv}$ to calculate the free energy. \markup{For each initial value of the $\Delta_i$s, we thereby obtain $f$ and $\Keffinv_{\qv}$ with a corresponding $T$.}

\markup{For a random initial self-energy the NBT might not converge to the lowest temperatures. } In those cases we initialize the iterations using a guessed form of $\Keffinv_{\qv}$ with peaks at suitable momenta. If different initial conditions converge to different states, we pick the state with the lowest free energy.


To get information about the spin correlations we calculate the quantity
\be
A_{\qv} \equiv \sum_{ij} \Keffinv_{\qv, ij} e^{-i \qv \cdot (\alpha_i -\alpha_j)}.
\ee
 $A_{\qv}$ is periodic with twice the reciprocal lattice vectors and the associated extended Brillouin zone (EBZ) is a truncated octahedron with dimensions twice those of the first Brillouin zone (1BZ) of the fcc lattice.
$A_{\qv}$ is closely related to the spin structure factor $S(\qv) \equiv \sum_{ij} \langle \Sv_{-\qv,i} \cdot \Sv_{\qv,j} \rangle e^{i \qv \left( \alpha_i -\alpha_j \right)}  =N_s T \left(A_{\qv} + A_{-\qv} \right)/4$. While $S(\qv)$ is manifestly inversion symmetric, $A_{\qv}$ is not as it reflects the symmetries of the self-energy $\Sigma_{\qv,ij}$. We take lack of inversion symmetry in $A_{\qv}$ to indicate that the spin state breaks inversion symmetry.
\\
\\

\section{Results \label{sec:results}}
\noindent
For the pure AF nearest-neighbor Hamiltonian, NBT gives no symmetry breaking down to the lowest temperature studied ($T \simeq 10^{-9}$), and $A_{\qv}$ shows $O_h$ symmetric extended maxima on the square surfaces of the EBZ with pinch points at $\mathrm{X_{EBZ}}$~\footnote{We use the same labeling of points in the EBZ as in Ref.~\onlinecite{Iqbal2019}}.

\begin{figure}[]
\includegraphics[width=\columnwidth]{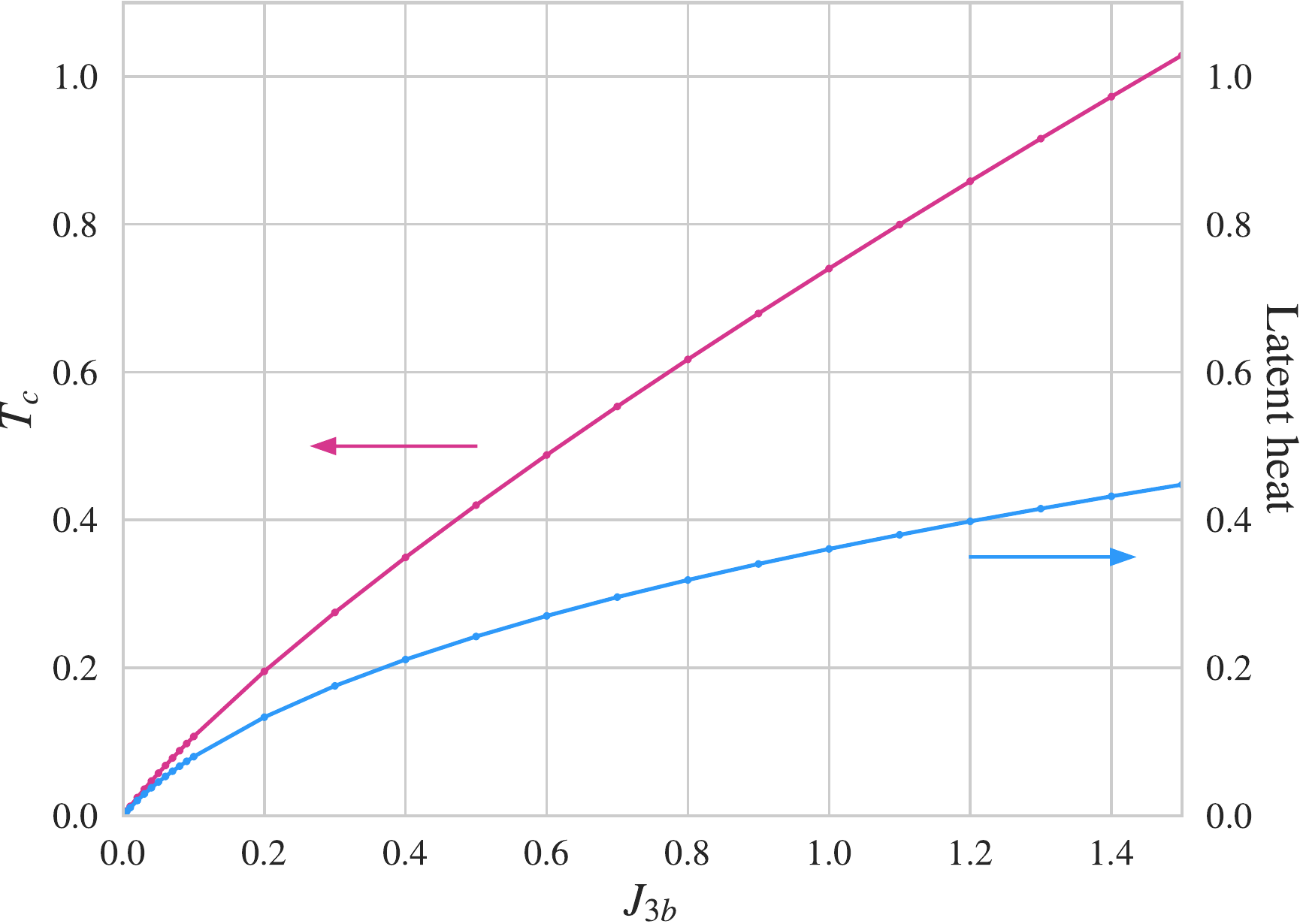}
\caption{Phase transition temperature $T_c$ \markup{(pink)} and latent heat \markup{(blue)} vs. $J_{3b}$ for the transition into the SLP state. \markup{$J_2 = J_{3a} = 0$}. $L=36$. \label{fig:TcvsJ3b} }
\end{figure}

For the $J_1$-$J_{3b}$ model with $J_{3b}=0.2$, the maxima of $A_{\qv}$ at high temperature occurs at $\mathrm{W_{EBZ}}$.  As the temperature is lowered, these maxima move into the hexagonal EBZ surfaces keeping the full $O_h$ symmetry. Then at $T_c =0.195$ (for $L=36$) the NBT free energy reveals a first-order phase transition into a low-temperature phase with a total of eight peaks in $A_{\qv}$ in the EBZ
\footnote{The specific symmetry breaking pattern depends on the random initial value of $\Sigma_{\qv}$. In all cases Eq.~\eqref{eq:down-tetrahedra-condition} is obeyed.}: at $\Qv_{(0,1)}+\bv_1 + n_2 \bv_2 + n_3 \bv_3$ for $n_2,n_3 \in \{0,1\}$, and at $\Qv_{(2,3)} + \bv_2 + n_1 \bv_1$, $\Qv_{(2,3)}+ \bv_3 + n_1 \bv_1$ for $n_1 \in \{0,1\}$, with $\Qv_{(0,1)} = (0,-4\pi/3, 4\pi/3)$ and $\Qv_{(2,3)} = (0,4\pi/3, 4\pi/3)$. $\bv_i$ denote the reciprocal lattice vectors for the fcc Bravais lattice. The symmetry of $A_{\qv}$ is thus reduced from $O_h$ to $C_{2v}$.
In particular, inversion symmetry and all three- and four-fold symmetries are broken. When adding also $A_{-\qv}$ to obtain the spin structure factor, the $C_{2v}$ symmetry is increased to $D_{2h}$. The peaks in $A_{\qv}$  can be explained as originating from a $120\degree$ SLP state where antiparallel spirals on sublattices 0 and 1 order at $\Qv_{(0,1)}$, and antiparallel spirals on sublattices 2 and 3 order at $\Qv_{(2,3)}$.
We find that such first-order transitions into the $C_{2v}$ symmetric $120\degree$ SLP phase occur for all positive values of $J_{3b}$, see Fig.~\ref{fig:TcvsJ3b}.

\begin{figure}[]
\includegraphics[width=\columnwidth]{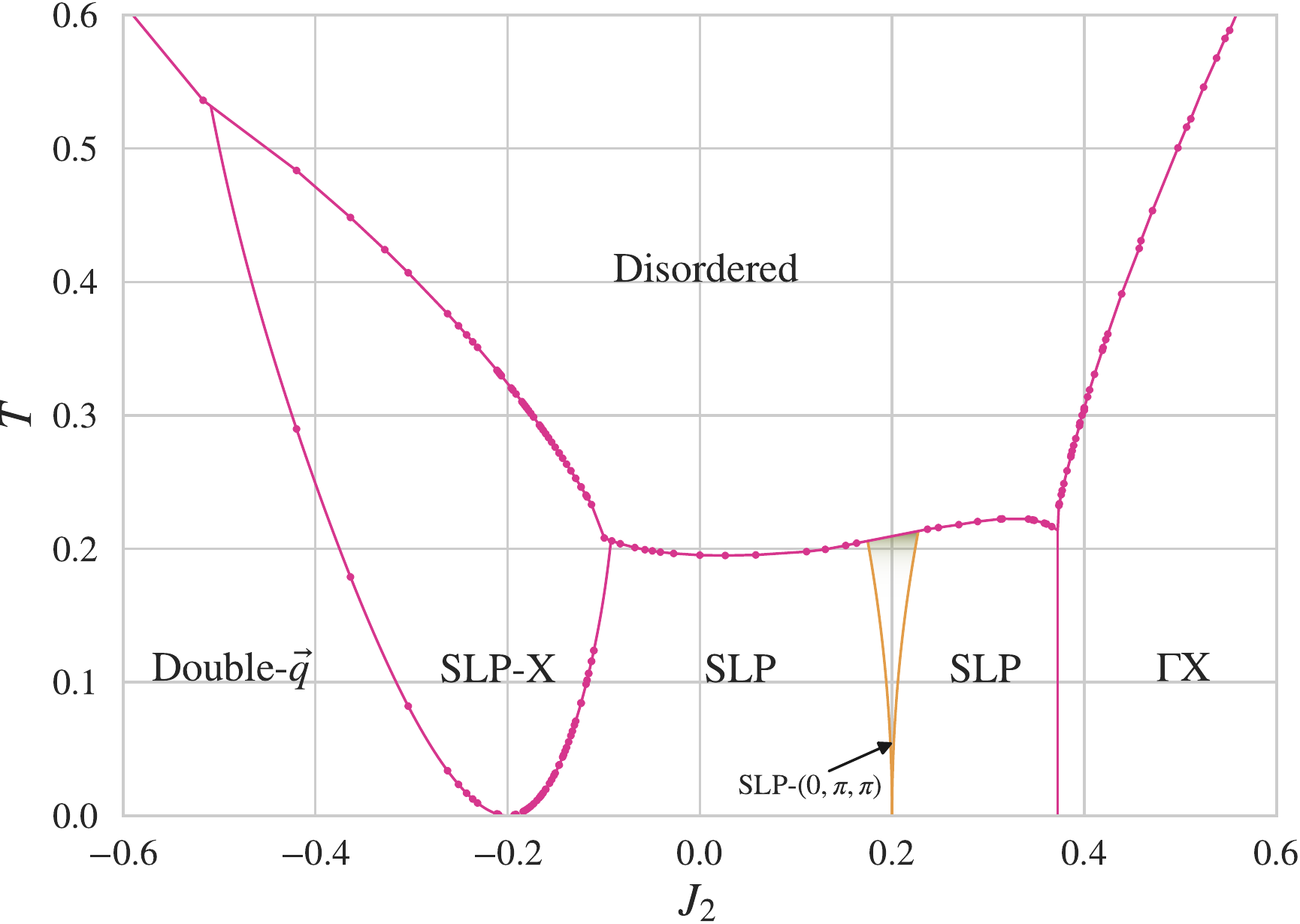}
\caption{$T$ vs. $J_2$ phase diagram for $J_{3b}=0.2$. Various system sizes $L=26-60$.  \markup{$J_{3a} = 0$}. The solid curves, based on discontinuities in the free energy (dots), are first-order phase transitions. The yellow lines around the SLP-$(0,\pi,\pi)$ phase are uncertain as NBT does not converge in the shaded region close to the disordered phase.}\label{fig:TcvsJ2}
\end{figure}
We next check the stability of the SLP phase when adding $J_2$. The finite-temperature phase diagram obtained using NBT for $J_{3b} = 0.2$ is shown in Fig. ~\ref{fig:TcvsJ2}.
The SLP phase is stable in the region $-0.2 \leq J_2 < 0.372$, and the ordering wave vectors follow Eq.~\eqref{eq:SLPQs}.
For FM $J_2 < -0.2$, the SLP state becomes unstable to a double-$\qv$ state, reminiscent of the multi-$\qv$ states investigated in Refs.~\onlinecite{Tsuneishi07,Chern08,Okubo2011,Lapa2012}, where two ordering vectors are present on \textit{all} sublattices.

For $J_2=-0.2$, we find a special case of the SLP state, labeled SLP-X in Fig.~\ref{fig:TcvsJ2}, where all sublattices have the same ordering vector $\mathrm{X}_{\mathrm{1BZ}}$. $A_{\qv}$ has maxima at opposite corners of four of the EBZ square surfaces;
$(2\pi,4\pi,0)$, $(2\pi,0,4\pi)$
, and subleading peaks with half maximum intensity at the four points $(0,\pm 2\pi,\pm 2\pi)$, consistent with the SLP ordering vector $\Qv_{(0,1)} = \Qv_{(2,3)} = (2\pi,0,0)$. The peak locations transform into each other by the $D_{4h}$ subgroup of $O_h$. Thus, this phase is inversion symmetric, as opposed to the general SLP phase. This SLP-X phase extends both into the double-$\qv$ and the general SLP regions at finite temperatures, and will therefore generally cause two ordering transitions as the temperature is lowered from the disordered phase, first one into the SLP-X phase and then another into the double-$\qv$ or general SLP phase at a lower temperature, see Fig.~\ref{fig:TcvsJ2}.

For $J_2 = J_{3b}$ at low temperatures, when biased into it, NBT converges to an SLP-$(0,\pi,\pi)$ state where $A_{\qv}$ also has $D_{4h}$ symmetry. In this state the two spirals each reduce to a collinear configuration, one with ordering wave vectors $\Qv_{(0,1)} = \pm(0,-\pi,\pi)$ and the other with $\Qv_{(2,3)} = \pm (0,\pi,\pi)$.  This state exists up to a finite temperature, but not all the way up to the disordered phase. We have not found the proper state at the intermediate temperatures, as we have not been able to get NBT to converge in the shaded region in Fig.~\ref{fig:TcvsJ2}. Nevertheless, we have indicated phase boundaries around the SLP-$(0,\pi,\pi)$ phase as it necessarily must be separated by phase transitions from the inversion symmetry-breaking SLP phase surrounding it.

For sufficiently strong AF $J_2$, the SLP phase gives way to a single-$\qv$ state with an ordering wave vector along the $\Gamma \mathrm{X}_{\mathrm{1BZ}}$ line. In this phase all sublattices order at the same wave vector, and the tetrahedra conditions are no longer satisfied. \markup{The phase boundary between the SLP phase and the $\Gamma \mathrm{X}_{\mathrm{1BZ}}$ phase is estimated to lie between $J_2 = 0.366$ and $J_2 = 0.374$ from our NBT calculations. The vertical line at $J_2 = 0.372$ is chosen from where the minimum of $J_{\vec{q}}$ changes character.}

\begin{figure}[]
\includegraphics[width=\columnwidth]{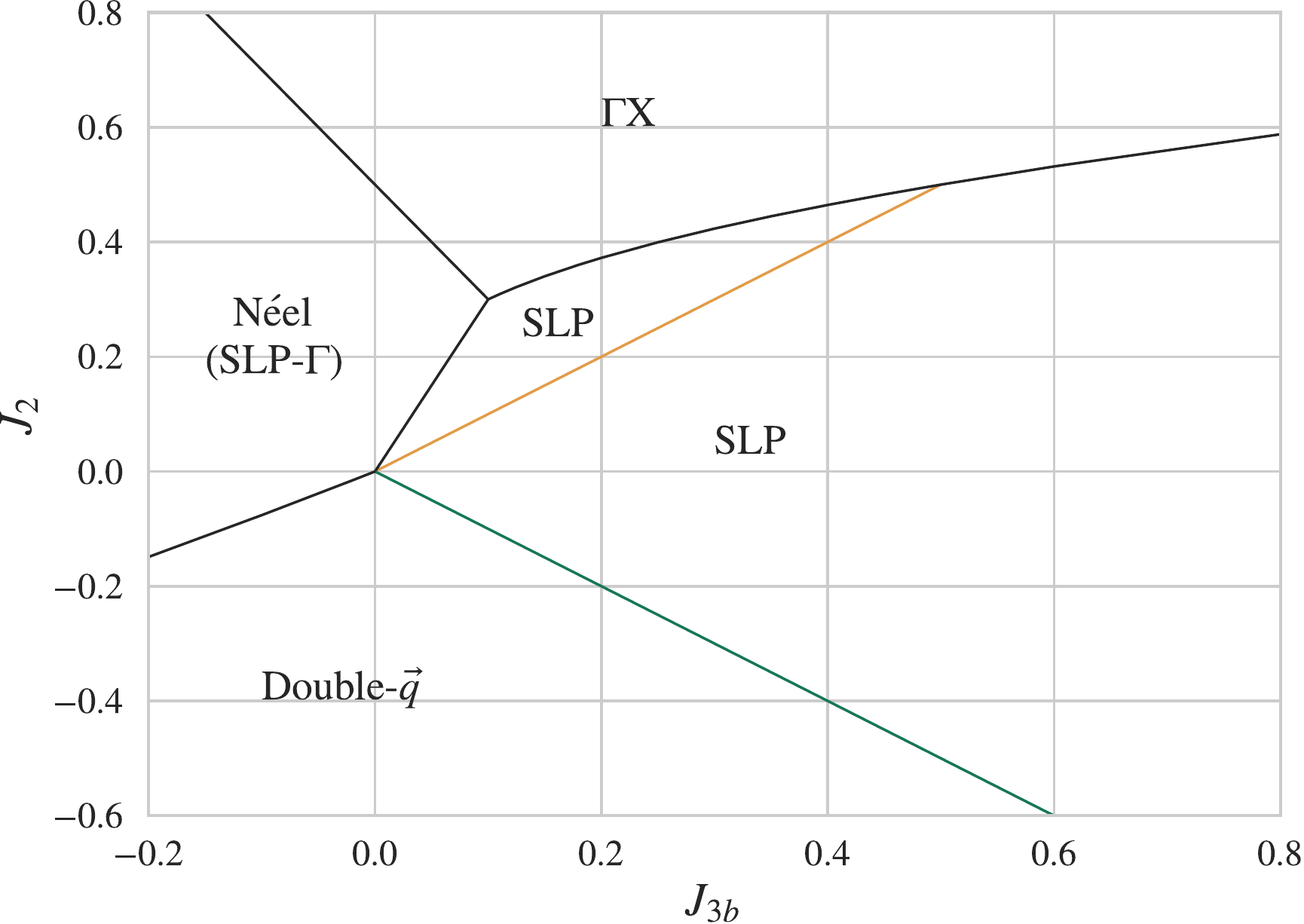}
\caption{Low-temperature phase diagram in the $J_{3b}$-$J_2$ coupling space, \markup{$J_{3a} = 0$}.  The inversion symmetric SLP-X and SLP-$(0,\pi,\pi)$ states are realized at $J_2=-J_{3b}$ (green line) and $J_2 = J_{3b}$ (yellow line), respectively. For large AF $J_{3b}$, the SLP state is stable for $-J_{3b} \leq J_2 < 1$.} \label{fig:Phasediagram}
\end{figure}

In Fig.~\ref{fig:Phasediagram} we map out the low-temperature phase diagram in the $J_{3b}$-$J_2$ coupling space using NBT. It is seen that the SLP phase exists in a large region.

On the AF $J_2$ side, the SLP phase ceases to exist when it becomes energetically favorable to violate the tetrahedra conditions. For small and intermediate values of $J_{3b}$ we find the single-$\qv$  $\Gamma \mathrm{X}_{\mathrm{1BZ}}$ phase. For large $J_{3b}$ the SLP phase is stable up to $J_2 = 1$, which is the limit set by the mapping from $J_1$-$J_2$ to $\tilde{J}_1$-$\tilde{J}_{3a}$.

On the FM $J_2$ side, the SLP phase borders the double-$\qv$ phase where \textit{each} sublattice has two ordering vectors. The two ordering vectors are $(2\pi,l,l)$ and $(2\pi,l,-l)$, where $l$ increases from $0$ ($\mathrm{X}_{1\mathrm{BZ}}$) to $\pi/2$ ($\mathrm{U}_{1\mathrm{BZ}}$) as $J_{3b}$ decreases from $-J_2$.  It reaches $\pi/2$ for $J_{3b}$ equal to a small positive $J_2$-dependent value. For $J_{3b}$ less than this, the two ordering vectors shift to $(0,k,k)$ and $(0,-k,k)$ with $k$ decreasing slowly from $3\pi/2$ ($\mathrm{K}_{1\mathrm{BZ}}$) as $J_{3b}$ decreases further. Spins in such double-${\qv}$ spirals will only obey the length constraint when the number of spin components $N_s>3$.  The NBT is derived from the large-$N_s$ limit without vertex corrections, and is extrapolated down to $N_s=3$. We believe that the double-${\qv}$ state produced by NBT is a remnant of the large-$N_s$ limit and that the extrapolation down to $N_s=3$ does not take the length constraint sufficiently into account.

We have also performed a similar stability analysis in $J_{3b}$-$J_{3a}$ coupling space with $J_2=0$. There, for AF $J_{3b}$, the SLP state is stable in an even wider region; whenever $J_{3a} \leq J_{3b}$.

To get information about the relative orientation of the spiral plane vectors $\vec{u}_{(i,j)}$ and $\vec{v}_{(i,j)}$ of the two SLP spirals, we have performed a spin wave calculation to compute the entropy.
We find that entropy favors the SLP state to be coplanar for our model, Eq.~\eqref{eq:Hamiltonian}, with the two SLP spirals sharing spiral plane vectors. Consequently, entropy favors collinear states in the special cases of SLP-$\Gamma$ and SLP-X.

\section{Discussion\label{sec:discussion}}
\noindent

We have shown how the classical AF Heisenberg model on the pyrochlore lattice with small further-neighbor couplings orders with coplanar SLP. In SLP, pairs of sublattices form antiparallel spirals. The ordering wave vectors are in general different for the two sublattice pairs, and are found to be the wave vectors that minimize the total $\tilde{J}_{3a}$-$J_{3b}$ energy of the paired fcc sublattices subject to the tetrahedra conditions.

For the pure $J_1$-$J_{3b}$ model, we find a $120\degree$ SLP state. This state simultaneously satisfies both AF $J_1$ and AF $J_{3b}$ and is thus realized for all $J_{3b} > 0$. It is separated from the disordered phase by a first-order phase transition with a $T_c$ and latent heat that goes to zero as $J_{3b} \to 0$, Fig.~\ref{fig:TcvsJ3b}. This is consistent with the AF nearest-neighbor model not ordering~\cite{Villain1979,Reimers91, Reimers92, Moessner98}.
Note that the critical temperatures are likely to be overestimated as NBT excludes vertex corrections~\cite{Syljuaasen2019}.

The SLP state generally breaks inversion symmetry, except when four times the ordering wave vectors are reciprocal lattice vectors.
We note that recent numerical results indicate that quantum fluctuations of the purely AF spin-1/2 and spin-1 models also induce inversion symmetry-breaking~\cite{Hagymasi21,Astrakhantsev21,Hagymasi22}.

As special cases of the SLP state, where all sublattices order at the same wave vector, we find SLP-$\Gamma$ (Néel) and SLP-X for our model. SLP-$\Gamma$ (Néel) has previously been identified as the ground state for the $J_1$-$J_2$ model with small AF $J_2$~\cite{Chern08, Lapa2012,Iqbal2019}. SLP-X is realized along the line $\tilde{J}_{3a} = J_{3b} > 0$ (and close to this line at intermediate temperatures) and should be the symmetry-broken state for the $J_1$-$J_3$ model in Ref.~\onlinecite{ConlonChalker2010}. \markup{An \textit{ab inito} study of the breathing pyrochlore material $\mathrm{LiGaCr}_4\mathrm{O}_8$ finds $\tilde{J}_{3a} = J_{3b} > 0$ and SLP-X as the corresponding low-temperature state~\cite{Ghosh19}.  They also find SLP-X to be stabilized at intermediate temperatures for $\mathrm{LiInCr}_4\mathrm{O}_8$, where $\tilde{J}_{3a} \approx 1.8 J_{3b} > 0$, which could be explained by the finite-temperature extension of the SLP-X phase due to its collinearity.}

The $J_{2}$ and AF $J_{3a}$ bonds favor states which do not satisfy the tetrahedra conditions. Nevertheless, we find that the SLP state dominates a large portion of the exchange coupling space, particularly in the region $J_1 > J_{3b} > \tilde{J}_{3a}$ for AF $J_1$ and $J_{3b}$.

This region might be relevant for the $\mathrm{Gd}_2B_2\mathrm{O}_7$ class of materials ($B$ is a nonmagnetic cation)~\cite{Wills2006}.  These are believed to be described as classical pyrochlore Heisenberg AFs with further-neighbor and dipole-dipole interactions~\cite{Raju1999,Ramirez2002,Welch2022}.
While we have not considered the dipole-dipole interaction, we note that a recent experiment on $\mathrm{Gd}_2\mathrm{Ti}_2\mathrm{O}_7$ suggests a partially ordered state with two ordering vectors at different $\mathrm{L}_{\mathrm{1BZ}}$ points~\cite{Paddison2021}. For SLP states such L-peaks are most likely to occur at intermediate temperatures near the line $J_2=J_{3b} >0$ where there are line minima in-between the $\mathrm{L}_{\mathrm{1BZ}}$ points.

The pyrochlore spinel materials are expected to have $J_{3a}$ at the same order of magnitude as $J_{3b}$~\cite{Yaresko08, Cheng08}.  As we find the SLP phase to be stable for $J_{3a} \leq J_{3b}$ for AF $J_{3b}$, it could be of relevance also for this class of materials. Especially in the ferrites $A\mathrm{Fe}_2\mathrm{O}_4$, the two types of third nearest-neighbor couplings have been suggested to be of comparable strength~\cite{Cheng08}. Inclusion of an AF $J_2$ could also help to push the system towards the SLP phase.

We envision also the \markup{general} SLP states to be realized on the breathing pyrochlore lattice when both the ``nearest"-neighbor couplings $J_1$ and $J_1^\prime$ are sufficiently strong AF, such that the tetrahedra conditions are satisfied. For further work it would be interesting to study the stability of the SLP state on the breathing pyrochlore lattice as well as its stability when adding dipolar interactions and/or anisotropy. \markup{We also note that the extension of NBT to the pyrochlore lattice can be used to study other interesting problems such as possible symmetry breaking in the $J_2 = J_{3a}$ model~\cite{Mizoguchi2018}.}

\begin{acknowledgments}
We would like to thank C. Castelnovo for pointing us in the direction of the $J_1$-$J_{3b}$ model, and J. Paaske and O.O.L.~Solow for useful discussions about NBT. C.G. acknowledges funding from the Aker Scholarship. The computations were performed on resources provided by
Sigma2 - the National Infrastructure for High Performance Computing and
Data Storage in Norway.
\end{acknowledgments}

\bibliography{p13.bbl}

\end{document}